\documentclass{naturefig}
\usepackage{hyperref}
\usepackage{amsmath}
\usepackage{amssymb}
\usepackage{graphicx}
\usepackage[utf8]{inputenc}
\usepackage[font={small
}]{caption}
\usepackage[a4paper,top=1.1in,bottom=1.1in,%
            footskip=.25in]{geometry}
\let\OLDthebibliography\thebibliography
\renewcommand\thebibliography[1]{
  \OLDthebibliography{#1}
  \setlength{\parskip}{0pt}
  \setlength{\itemsep}{0pt plus 0.3ex}
}

\title{How Dark Matter Came to Matter
}

\author{Jaco de Swart,$^{1,2,3}$ Gianfranco Bertone$^{1,3}$ \& Jeroen van Dongen$^{1,2}$}

\begin{document}

\maketitle
\vspace{18pt}
\begin{affiliations}
 \item Institute for Theoretical Physics Amsterdam, University of Amsterdam, Science Park 904, 1098 XH Amsterdam, The Netherlands
 \item Vossius Center for the History of the Humanities and Sciences, University of Amsterdam, Science Park 904, 1098 XH Amsterdam, The Netherlands
 \item GRAPPA, University of Amsterdam, Science Park 904, 1098 XH Amsterdam, The Netherlands
\end{affiliations}


\begin{abstract}
The history of the dark matter problem can be traced back to at least the 1930s, but it was not until the early 1970s that the issue of `missing matter' was widely recognized as problematic. In the latter period, previously separate issues involving missing mass were brought together in a single anomaly. We argue that reference to a straightforward `accumulation of evidence' alone is inadequate to comprehend this episode.
Rather, the rise of cosmological research, the accompanying renewed interest in the theory of relativity and changes in the manpower division of astronomy in the 1960s are key to understanding how dark matter came to matter. At the same time, this story may also enlighten us on the methodological dimensions of past practices of physics and cosmology.
\end{abstract}

\vspace{3ex}

\noindent
Uncovering the nature of `dark matter'---the mysterious substance that dominates the mass budget of the universe from sub-galactic to cosmological scales---is arguably one of the greatest challenges of modern physics and cosmology. Indeed, many physicists and astronomers across the world are today trying to identify the nature of dark matter.\cite{Bertone2010} In 2016 alone, for example, an average of at least three publications with ``dark matter'' in their title appeared every day (see the SAO/NASA Astrophysical Data System).

The dark matter problem has a surprisingly long history.\cite{Trimble1987,Trimble2013,Sanders2010,Einasto2014,Bertone2016} Already in the early 20th century physicists and astronomers attempted to estimate the amount of non-lumin\-ous matter in the Galaxy.\cite{Kapteyn1922,Jeans1922,Oort1927} In the 1930s several authors noticed an inconsistency between the observed velocity dispersion of galaxies in galaxy clusters and that same dispersion as it followed from calculations on the basis of visible, luminous matter.\cite{Zwicky1933,Smith1936a,Holmberg1937} The problematic nature of these early observations, however, did not become a central concern until much later, in the early 1970s, when astronomers reinterpreted an inconsistency between the observed `flat' rotation curves of gas in galaxies and the `declining' curves that had been predicted on the basis of the observed stars in those systems.\cite{Faber1979}

The eventual acceptance of an hypothesis of `dark matter' is often understood as an example of the accumulation of unequivocal evidence: two results from different branches of astronomy---high velocity dispersions in clusters and flat rotation curves in galaxies---would have indicated unexpectedly large galaxy masses, and only by the early 1970s had enough evidence accumulated to accept the existence of a preponderance of yet unobserved matter.\cite{Riess2015} Such an account captures the broad scope of the research involved. Nevertheless, it fails to clarify \emph{why} astronomers started to see two independent results as being due to a single problem of missing mass in the first place.

In order to understand how dark matter came to matter, we need to understand how and why the mutually independent and problematic observations acquired the status of \emph{evidence}, and the status of \emph{evidence for the hypothesis of missing mass} in particular. It will be important to look at the conceptual and institutional changes in astronomy during the 1960s and early 1970s: these directed astronomers towards cosmology. Subsequently, as we will see, the discrepancies observed in both rotation curves and galaxy clusters were brought together in the search for the cosmological mass density of the universe, leading to the  publication of two key papers in 1974 that argued for the existence of additional, unobserved matter.\cite{Ostriker1974,Einasto1974} In the wake of these articles dark matter grew to be seen as an anomaly, and eventually transformed into a central concept of the current paradigm of cosmology. Finally, one may ask whether the story of dark matter also has a larger significance: as this case study informs us about past practices of physics and cosmology, can it also inform us in current debates on the proper methodology of the field?

\vspace{8pt}

\section*{Mass Discrepancy in Clusters of Galaxies}

Our story begins in 1933, when Swiss-born astronomer Fritz Zwicky famously published a paper on the Coma cluster of galaxies.\cite{Zwicky1933} Zwicky found the velocity dispersion of its members to be so high that, to keep the system stable, the average mass density in the Coma system would have to be much higher than that deduced from observed visible matter. He attributed this to the presence of yet unseen, dark matter. Similar results were found soon afterwards by Swedish astronomer Erik Holmberg, who studied systems of galaxies, and American physicist Sinclair Smith, who analysed the mass of the Virgo cluster.\cite{Holmberg1937,Smith1936a} The latter, too, noted that ``a discrepancy appears'' 
between the two ways of determining the masses of galaxies.

Despite the discrepancy, debate on these observations only began after new astronomical surveys and catalogues had sparked an interest in clusters of galaxies in the late 1950s.\cite{Zwicky1956a,Shane1956,Abell1957,Abell1959} Their masses were now studied in more detail, and different systems of galaxies were found in which the mass discrepancy again appeared.\cite{Kahn1959,Page1959} Prompted by these developments, Soviet-Armenian astrophysicist Viktor Ambartsumian proposed a new explanation for the awkward observations in 1958, an explanation that was different from Zwicky's hypothesis of unknown dark matter. Ambartsumian argued that the observed discrepancies were due to the absence of `dynamical equilibrium' in these groups and clusters: their galaxies were actually rapidly flying apart, which was to produce the unusual data.\cite{Ambartsumian1958a}

Ambartsumian's controversial idea quickly became an influential hypothesis in the analysis of clusters.\cite{Burbidge1959} In 1961, a conference dedicated to critically examine Ambartsumian's instability hypothesis took place in Santa Barbara, CA. Here, the idea of cluster instability as well as that of unseen mass were vividly discussed, together with other less popular alternatives. Both the hypothesis of instability and of additional matter were problematic. If groups and clusters of galaxies were indeed unstable, then they would not last more than 10 to 1000 million years, which was argued to be very short compared to the time scale of the universe.\cite{VandenBergh1961} Most clusters should already have been dissolved in such a scenario, in conflict with observation. The alternative, however, was equally ``distasteful'', 
according to the conference organizers: this would imply that in the field of astronomy ``theories are based on  observations of less than 1\% of the matter that is really there!''\cite{Neyman1961a} The conference instilled a sense of urgency regarding these problems in its participants. At the same time, the contributed presentations reflected a strong difference of opinion on how to best account for the discrepant observations.

After the event on instability, Santa Barbara hosted ``Problems of Extragalactic Research", the fifteenth symposium of the International Astronomical Union. Influential astronomer Geoffrey Burbidge here reported that it was down to a ``matter of taste'' which account of clusters one considered to be correct.\cite{Burbidge1962} 
Indeed, well into the late 1960s and early 1970s many different solutions to the cluster discrepancies were discussed. These included ideas about possible regions of ionized hydrogen;\cite{Woolf1967} the existence of a large number of dwarf galaxies;\cite{Reddish1968} changes to the law of gravity;\cite{Finzi1963,Forman1970,Jackson1970} a large density of gravitational radiation;\cite{Field1971} cosmologically created black holes;\cite{Hawking1971} the notion that separate field galaxies could have been mistaken for cluster members;\cite{Gott1973} the presence of massive neutrinos;\cite{Cowsik1972} or, finally, the possibility of observational errors.\cite{Abell1975} Whether clusters were unstable was still heavily debated as well.\cite{Karachentsev1966,Burbidge1969,DeVaucouleurs1975} Too few observational and theoretical constraints were available to force a consensus on how to interpret the discrepancy. The existence of additional, unobserved mass was just one possibility among a considerable number of alternatives.

\vspace{8pt}

\section*{Flat Galactic Rotation Curves}

`Rotation curves' are diagrams representing the orbital velocity of gas and stars in galaxies as function of their distance to the galactic centre. They are tools to study the kinematics of galaxies, and provide a way to estimate their masses. The first rotation curves were produced for nearby galaxies like M31, M33, or the Milky Way itself.\cite{Lindblad1927,Oort1927,Babcock1939,Mayall1942} The computation of rotation curves became increasingly prominent after the discovery of the 21-cm radio emission line\cite{Ewen1951,Muller1951,Pawsey1951} (of otherwise invisible neutral hydrogen) and the rise of radio astronomy in the 1950s and 1960s. During the early 1970s a specific feature of these curves was coming to light: rotation curves of galaxies tend to be `flat'.

\begin{figure}[t]
\centering
\includegraphics[width=0.85\textwidth]{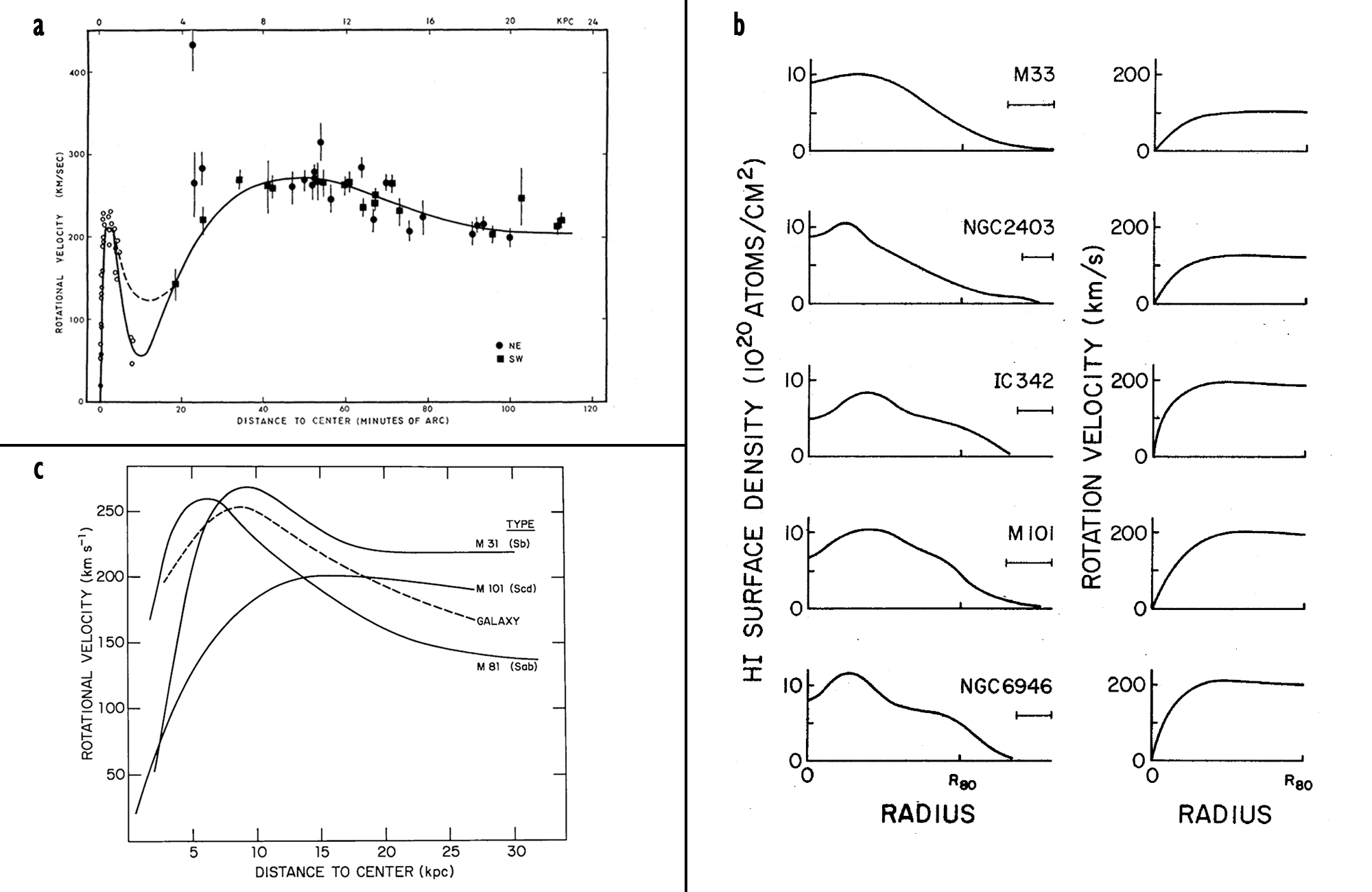}
\caption{\label{fig:rotcurv}Indications of flat rotation curves in the early 1970s. \textbf{a}, optically studied rotation of ionized hydrogen in M31 by Rubin and Ford (1970). \textbf{b}, five rotation curves of neutral hydrogen in different galaxies by Rogstad and Shostak (1972); R80 indicates the radius within which 80 percent of the neutral hydrogen mass is contained. \textbf{c}, rotation curves of three galaxies and the Milky Way by Roberts and Rots (1973). Reproduced from ref. 48, AAS/IOP (a); ref. 51, AAS/IOP (b); and ref. 49, EDP Sciences (c).}
\end{figure}

Flat rotation curves were a surprising find. The velocity of a rotating disk of stars and gas is expected to decline beyond the radial distance to which most mass is interior---a feature referred to as `Keplerian', in reference to Kepler's familiar laws that describe the orbital velocities in our solar system. If, instead, the rotational velocity stays constant (that is, flat) as the radius increases, then this indicates that there is more gravity than expected on the basis of the galaxy's observed light. This is the problem of `flatness' which today is often pointed out to argue that dark matter should exist.
Even though there were plenty of observations of flat rotation curves in the early 1970s (see Fig. \ref{fig:rotcurv}), interpretations of their consequences for the existence of unseen mass were scarce and lacked urgency.

Two important studies of rotation curves were published in 1970. Kenneth Freeman, and, separately, Vera Rubin and Kent Ford had studied galactic rotational velocities in the optical waveband. Freeman found that for two galaxies (NGC 300 and M33) the observed velocity maxima occurred at a much larger radius than predicted on the basis of stellar photometry. He mentioned the point in the appendix of his paper, where he noted that this could imply the existence of ``additional matter which is undetected,''\cite{Freeman1970} without further elaborating on the issue. Rubin and Ford analysed the rotation of the Andromeda Nebula out to large radii. Famously, they observed that the velocities stayed rather constant with radius (Fig. \ref{fig:rotcurv}a). Still, Rubin and Ford drew no direct conclusions regarding the existence of any dark matter or extra mass on the basis of their measurements---they only argued that how one may wish to extrapolate the curve beyond the furthest out measured velocity point was simply a ``matter of taste''.\cite{Rubin1970} 

Substantial numbers of radio astronomical analyses of the rotation of galaxies also started to appear in the early 1970s. Three galaxies with rotation curves that hinted at the possible presence of undetected mass at large radii were found through joint work of Morton Roberts, of the American National Radio Astronomical Observatory (NRAO) in Greenbank, WV, and Arnold Rots, of the Dutch Westerbork Synthesis Radio Telescope (WSRT) (Fig. \ref{fig:rotcurv}c). Roberts and Rots expressed that their find need not have been too surprising, as there was no evidence to favour ``small'' over ``large'' galaxies;\cite{Roberts1973} 
in other words, there existed no reason to assume that the ratio of mass to luminosity would be constant with galactic radius: indeed, there may be matter that is not visible. Yet, they, too, did not argue their point with any great urgency or emphasis: the reader was not suggested to be witness to a crucial discovery on the content and mass of galaxies. Similarly, Seth Shostak and Dave Rogstad of Owens Valley Radio Observatory evaluated spiral galaxies to find extended flatness in their rotation curves (Fig. \ref{fig:rotcurv}b);\cite{Shostak1972,Rogstad1972} they concluded that ``any extrapolation of the total mass [...] is very uncertain''.\cite{Rogstad1973} 

As the above examples have suggested, how flat rotation curves were to be extrapolated and what those extrapolations implied was uncertain and arguably considered a matter of taste---quite similar to how the mass discrepancy in galaxy clusters was appreciated. Moreover, there were many Keplerian models for galaxies available,\cite{Brandt1960,Toomre1963} and depending on how one played with the parameters, these could still be fitted to the extended curves.\cite{Rogstad1972,Huchtmeier1975} Indeed, some studies arrived at quite different results: a Cambridge group of radio astronomers found a rotation curve for M31 that perfectly matched the expected decreasing curve.\cite{Emerson1973,Baldwin1975} This, of course, was in flagrant contradiction to the problematic flatness inferred by others.\cite{Roberts1975}

In toto, then, the possible existence of unseen mass was a \textit{potential} solution to two independent problems that arose in the 1960s and early 1970s. The suggestion was highly uncertain and itself problematic, if considered at all. Indeed, there was no consensus, in either case, on what a proper interpretation of observed results should be, and there was no definite sense as to how much weight might be attributed to any interpretation. Clearly, in neither branch of astronomy were these observations at this point cited as as positive \textit{evidence} for the presence of extra matter, or falsifying evidence for any alternative hypothesis. Furthermore (besides a few exceptions)\cite{Schwarzschild1954,Finzi1963} the two problems were studied separately.

Flat rotation curves and galactic velocity dispersions only became evidence for unobserved mass after they were transferred to a new and different domain of research. This happened after substantial changes to both the research interests and institutional context of astronomers had taken place.

\vspace{8pt}

\section*{Changing Subjects of Interest and the Rise of Cosmology}

During the 1960s astronomical research involved increasingly larger scales and higher energies. While in the early twentieth century it was still generally believed that our galaxy constituted the entire universe, by midcentury this conception had changed dramatically. Astronomy had moved into extra-galactic studies involving the evolution of the entire universe,\cite{Smith2008a,Smith2009} while radio astronomy had opened an immense new window of wavelengths, which produced important breakthroughs.\cite{Roberts1965} Most prominent was the optical identification of quasi-stellar radio sources (QSRS or `quasars') in 1963.
The first identified quasar was seen to recede from us at a stunning velocity of 16\% of the speed of light ($z=0.158$) and was estimated to be optically a hundred times brighter than any other known radio object.\cite{Schmidt1963} These values, together with radio source counts and the observation of the microwave background radiation a year later, contributed substantially to the demise of the steady state theory of the universe.\cite{Kragh2006}

Partly in response to the enormous energy release indicated by the quasar discovery, the first of the successful `Texas Symposia on Relativistic Astrophysics' was quickly set up. The symposium marks the arrival of an era in which the boundaries between physics and astronomy became permanently blurred. In particular, general relativity, which earlier had been relegated to the quiet pace of mathematics and Albert Einstein's increasingly unlikely unification attempts,\cite{Eisenstaedt2006,Dongen2010} was climbing out of the doldrums and now quickly became the focus of attention. ``Everyone is pleased,'' Cornell's Thomas Gold orated at the conference dinner in Dallas: ``the relativists [...] who are suddenly experts in a field they hardly knew existed [and] the astrophysicists for having enlarged their domain, their empire, by the annexation of another subject---general relativity.''\cite{Gold1965} Historians have pointed out that this development capped the ``renaissance of general relativity.''\cite{Blum2015}

Indeed, the discovery of quasars sparked a new strong interest in cosmology. Their observed number versus redshift suggested that the universe was vastly different in the past than it is today,\cite{Sciama1971a} which produced urgent questions regarding the formation and evolution of galaxies.\cite{Page1964} Distances determined on the basis of the appearance of galaxies turned out to be unreliable, as the latter's shape, mass, brightness and other parameters were found to change rapidly within their lifetime.\cite{Tinsley1968a} Furthermore, the observation of pulsars in 1968 motivated theorists to try to further understand the mechanisms for small scale aggregation of mass\cite{Hewish1968}---these pulsars were soon identified as neutron stars.

``[T]he rapid pace of discovery in astronomy and astrophysics during the last few years has given this field an excitement unsurpassed in any other area of the physical sciences'' the National Science Board reported to the U.S. Congress in 1970.\cite{NationalResearchCouncil1972} 
As extragalactic phenomena, cosmology and the theory of general relativity were brought into immediate focus, the mass values of galaxies would become of the utmost importance for the practice of astronomy.

\vspace{8pt}


\section*{A Discipline in Flux}

\begin{figure}[t]
\centering
\includegraphics[width=0.41\textwidth]{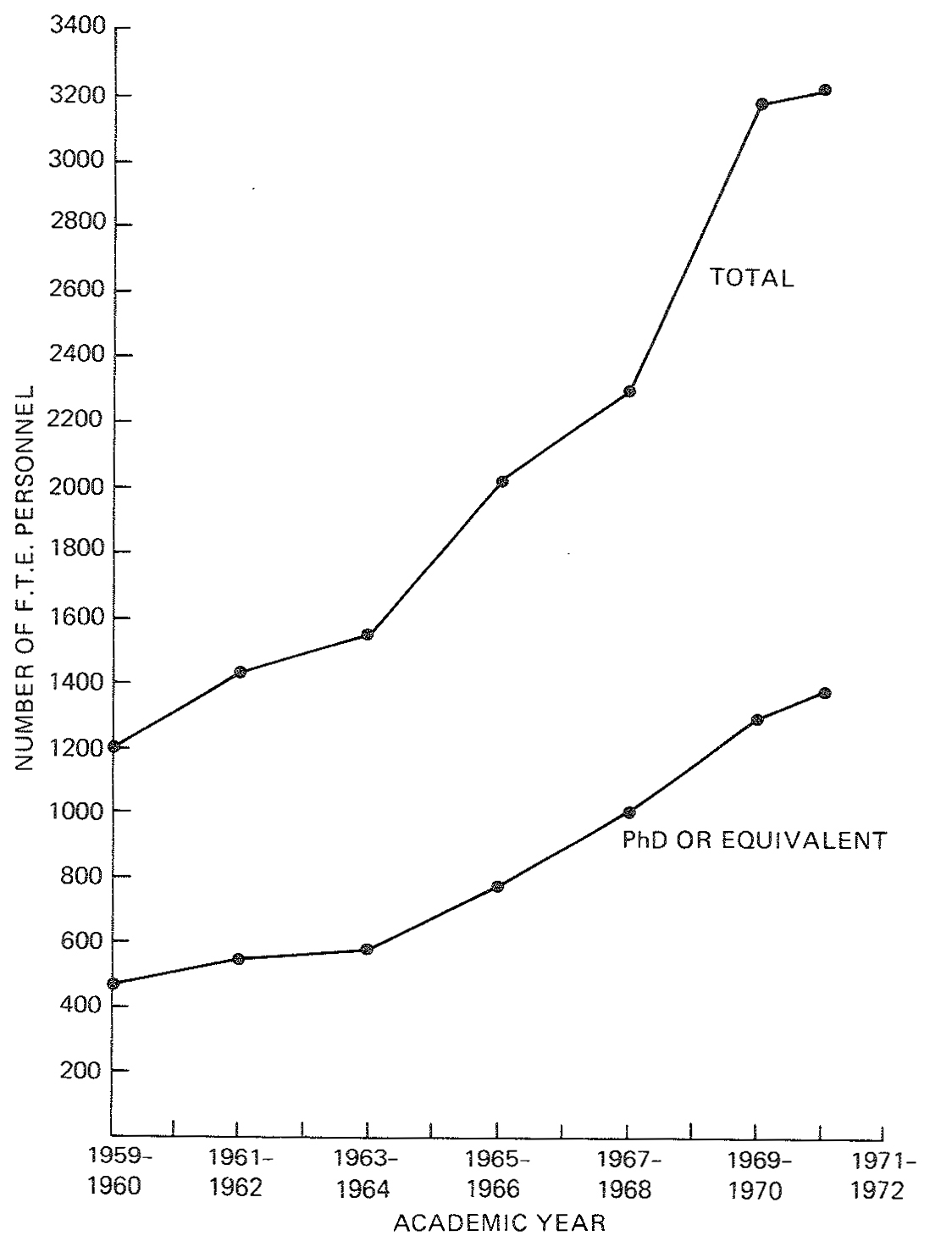}
\caption[Personnel in Astronomy, 1960-1970]{\label{fig:PhD2}The number of full-time equivalent (FTE) scientific and technical personnel employed in astronomy in the United States between the academic years 1959/1960 and 1969/1970. Source: National Research Council, (1972), \textit{Astronomy and Astrophysics for the 1970s: Volume 1: Report of the Astronomy Survey Committee}, National Academies Press, p. 57.}
    \end{figure}

The rise of cosmology took place during important changes in the institutions of astronomy and their population. Partly due to the demands of the American space program, there had already been a heavily increasing demand for astronomy graduates in the US:\cite{NationalResearchCouncil1964} between 1960 and 1970, its number of PhD granting astronomy departments more than doubled, faculty head count nearly tripled (see Fig. \ref{fig:PhD2}) and the number of degrees awarded increased tenfold.\cite{NationalResearchCouncil1973}

As disciplinary boundaries faded in cosmological research, departmental border crossings went up. This development was helped by the circumstance that American physicists were having a harder time finding work later in the decade: the Cold War d{\'e}tente that began in the late 1960s had led to a drop in defense funding for their field.\cite{Kaiser2002,Kaiser2006}  A large number of physicists now entered astronomy departments.  In 1966, 26 percent of the astronomy personnel with PhDs had received their doctorate in physics; within four years, this was at 45 percent.\cite{NationalResearchCouncil1973} The US National Research Council projected that it would take only two more years until there were more people working in astronomy with a physics PhD than an astronomy PhD.\cite{NationalResearchCouncil1972,NationalResearchCouncil1973}

The cosmological turn, however, also contributed to the creation of new job opportunities for astronomers at physics institutes: the number of astronomy graduates working in physics or combined physics and astronomy departments tripled between 1960 and 1970.\cite{NationalResearchCouncil1983} Finally, an increasing number of astronomy PhDs would indeed be awarded by `physics' or  joint `physics and astronomy' departments: from 8\% in 1960, to about one third in 1975.\cite{NationalResearchCouncil1983} 
Clearly, physics and astronomy had grown ever closer during the period, both in subject matter and institutionally.

\vspace{8pt}


\section*{How Cosmology gained its Weight}

\begin{figure}
\centering
\includegraphics[width=0.8\textwidth]{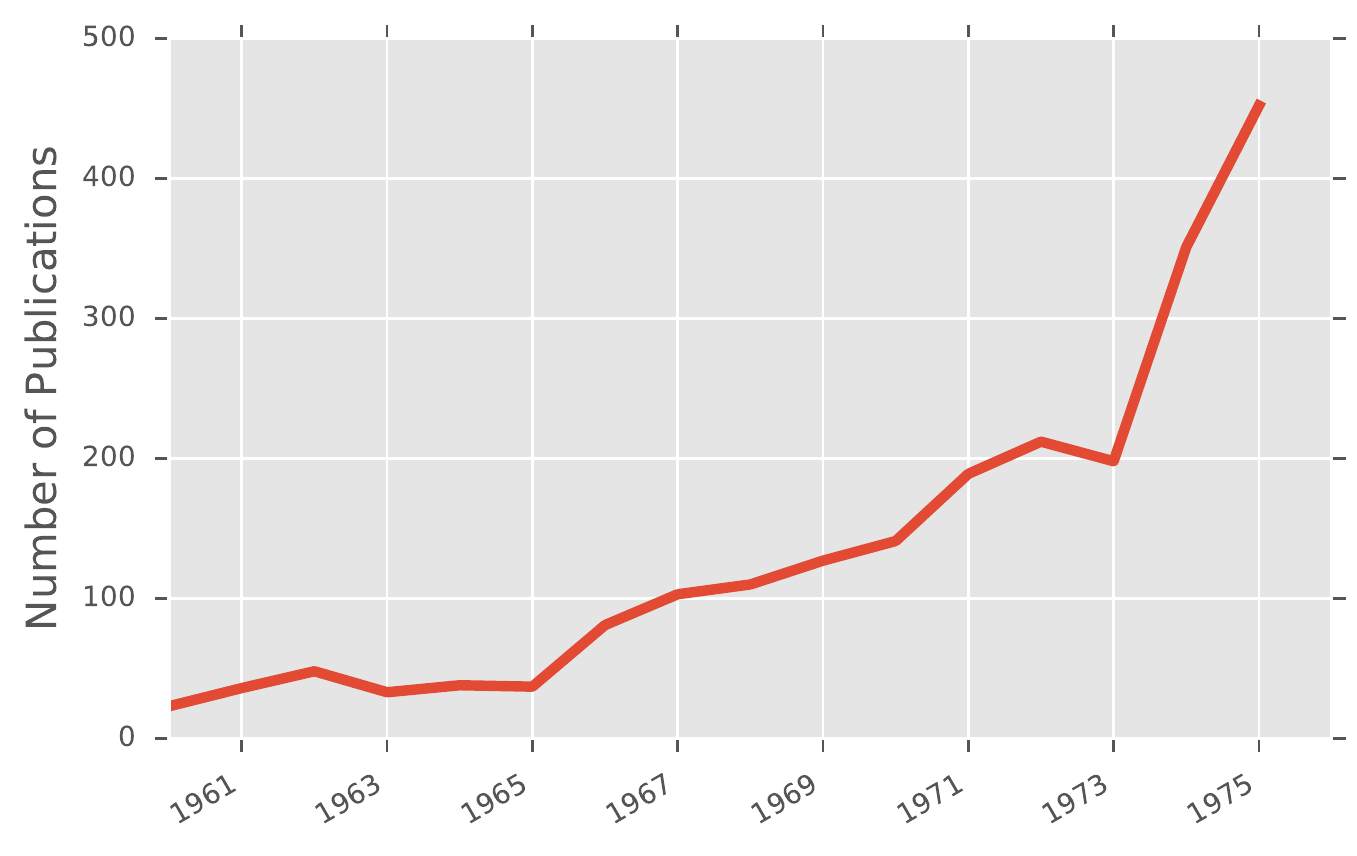}
\caption{\label{fig:cosm}
Number of publications between 1960 and 1975 with `cosmology' mentioned in abstracts or as keyword of publications in the SAO/NASA Astrophysical Data System.}
\end{figure}

The border crossings between physics and astronomy were productive: between 1965 and 1975 the number of refereed publications on cosmology increased tenfold (see Fig. \ref{fig:cosm}), and the subject's breakthroughs were quickly documented in textbooks whose influence would last for decades.\cite{Peebles1971,Sciama1971a,Weinberg1972,Misner1973,Hawking1973} Cosmology had become a quantitative subject, served by a plethora of observations, as reflected in its now often accompanying adjective  ``physical".\cite{Peebles1971,Smith2008a}

According to relativistic cosmology, a homogeneous and isotropic expanding universe can either be \textit{flat}, \textit{open}, or \textit{closed}, all corresponding to different fates of cosmic evolution. How to distinguish between these alternative scenarios of the universe?  The famed Friedmann equations that describe these scenarios
exhibit a struggle between cosmic expansion and gravity's attraction;  knowing which scenario is right, equals to knowing whether the universe would eventually collapse under gravity, or expand forever. The correct choice can be selected by determining the values of the parameters in the equations---this was, and still is, in principle a question of observation.

In 1970, Caltech astronomer Allan Sandage characterized this new key cosmological question as ``a search for two numbers'':\cite{Sandage1970}  the Hubble constant, $H_0$ (which sets the time scale of the universe) and the deceleration parameter, $q_0$. After a period of making readjustments, scholars had been nearing a consensus on the value of the Hubble constant.\cite{Trimble1996} However, the value of the deceleration parameter, which determines how much the universe's expansion is stagnating, was still uncertain. Many agreed that Einstein's cosmological constant was zero ($\Lambda=0$), and in the absence of a cosmological constant, the deceleration parameter would be fully determined by another parameter: the mass density of the universe, $\rho$.

Thus, in the early 1970s observing the mass density of the universe became of vital importance to the new and highly fashionable discipline of cosmology. The amount of matter in the universe was ``[o]ne of the most important quantities'' to detect and understand, Geoffrey Burbidge stated in 1970.\cite{Burbidge1972} Similarly, Dennis Sciama emphasized the ``key role'' of knowing the overall mean density of the universe in his 1971 textbook.\cite{Sciama1971a}

Yet, not only direct observations drove subsequent developments. A strong preference for a specific value of the universe's mass density existed, based on an \emph{a priori} conviction related to interpretations of Einstein's relativity theory. For example, in 1967 mathematical physicist Wolfgang Rindler noted that  ``\textit{[p]hilosoph\-ically}'',  a \textit{closed} universe  seemed ``most attractive''. It agreed  with Mach's principle, which entails that local inertial frames should be fully determined by all the matter in the universe, Rindler argued.\cite{Rindler1967} In a recent oral history interview, cosmologist James E. Gunn recalled that this ``aesthetic'' or ``religious'' view was shared among many theoretical cosmologists (J. E. Gunn, oral history interview with J.G. de Swart, 12 December 2014). For the universe to be closed, so that gravity overcomes the expansion due to the big bang, its mass density should be equal to or larger than a critical value of $\rho_c  \sim 10^{-29}~g/cm^{-3}$. The desire to `close the universe' motivated many estimates and calculations of local mass densities.\cite{Peebles1967,GottJ.R.1974}

In fact, many estimates of the average masses of galaxies already existed. These had  been produced by considering familiar  luminous matter and typically yielded a mass density of the universe of the order of $\sim 10^{-31} g/cm^{-3}$.\cite{Peebles1971,Shapiro1971,Noonan1971,Weinberg1972,Burbidge1972} This value was two orders of magnitude lower than the critical density necessary to close the universe. So, the visible mass density ($\rho$) of galaxies could be only a small fraction of the density needed to close the universe: $\Omega = \rho/\rho_c \sim 0.01$. With the new cosmological focus and its preferences, existing galaxy mass estimates produced a problematic discrepancy that quickly gave rise to ``much speculation'', as Steven Weinberg put it in 1972.\cite{Weinberg1972} 
In his words:
\begin{quote}
[I]f one tentatively accepts the result that $q_0$ is of order unity [$\Omega \geq 1$], then one is forced to the conclusion that the mass density of about $2 \times 10^{-29}g/cm^3$ must be found somewhere outside the normal galaxies. But where?\cite{Weinberg1972}
\end{quote}

\noindent So, the exciting new interdisciplinary subject of physical cosmology suddenly needed extra mass, on the basis of primarily a priori considerations. From this novel perspective, physicists and astronomers began to ponder about the universe's matter budget.

\vspace{8pt}

\section*{The Birth of an Anomaly}

At this point, only after astronomy had acquired a new cosmological focus that produced a search for extra matter, did the flat rotation curves and the cluster mass discrepancy come together. This took place in 1974, with the publication of two landmark papers. One was the product of a collaboration between physicist James Peebles and astronomers Jeremiah Ostriker and Amos Yahil at Princeton University.
Following earlier numerical work that had already suggested unseen mass in galaxy outskirts,\cite{Ostriker1973} they set out to estimate the total mass of galaxies (J. Ostriker, oral history interview with J.G. de Swart, 1 November 2014). Their paper was aptly titled ``The Size and Mass of Galaxies, and the Mass the Universe'' and it summarized existing derivations of the masses of galaxies and systems of galaxies to determine the average mass of the universe. The opening paragraph has become more than familiar to many: ``There are reasons, increasing in number and quality, to believe that the masses of ordinary galaxies may have been underestimated by a factor of 10 or more. [...] If we increase the mass of each galaxy by a factor well in excess of 10, we [...] conclude that observations may be consistent with a universe which is `just closed' ($\Omega=1$).''\cite{Ostriker1974}

When summing up the reasons for belief in unseen mass, Ostriker, Peebles and Yahil now used both the results on the mass discrepancy of groups and clusters,\cite{Page1961,Field1971,Rood1972} and data of flat rotation curves.\cite{Rogstad1972,Roberts1973} Ostriker \emph{et al.} concluded that galaxy masses increased proportionally with radius ($M(r) \propto r$) from 20 to 500 kpc; they argued that the high mass-to-light ratios and large galaxy sizes were best explained by a ``giant halo of faint stars''. 
In this scenario, galaxies accounted for at least one-fifth of the critical density, $\Omega_{galaxies} \geq 0.2$. This value was sufficiently close to $\Omega=1$ to suggest agreement with a closed universe, the authors implied. This somewhat generous extrapolation by a factor of five is suggestive of the desirability of that cosmological scenario, which was "believed strongly by some", the authors argued, "for essentially nonexperimental reasons".\cite{Ostriker1974}

\begin{figure}[t]
	\centering
	\includegraphics[width=1\textwidth]{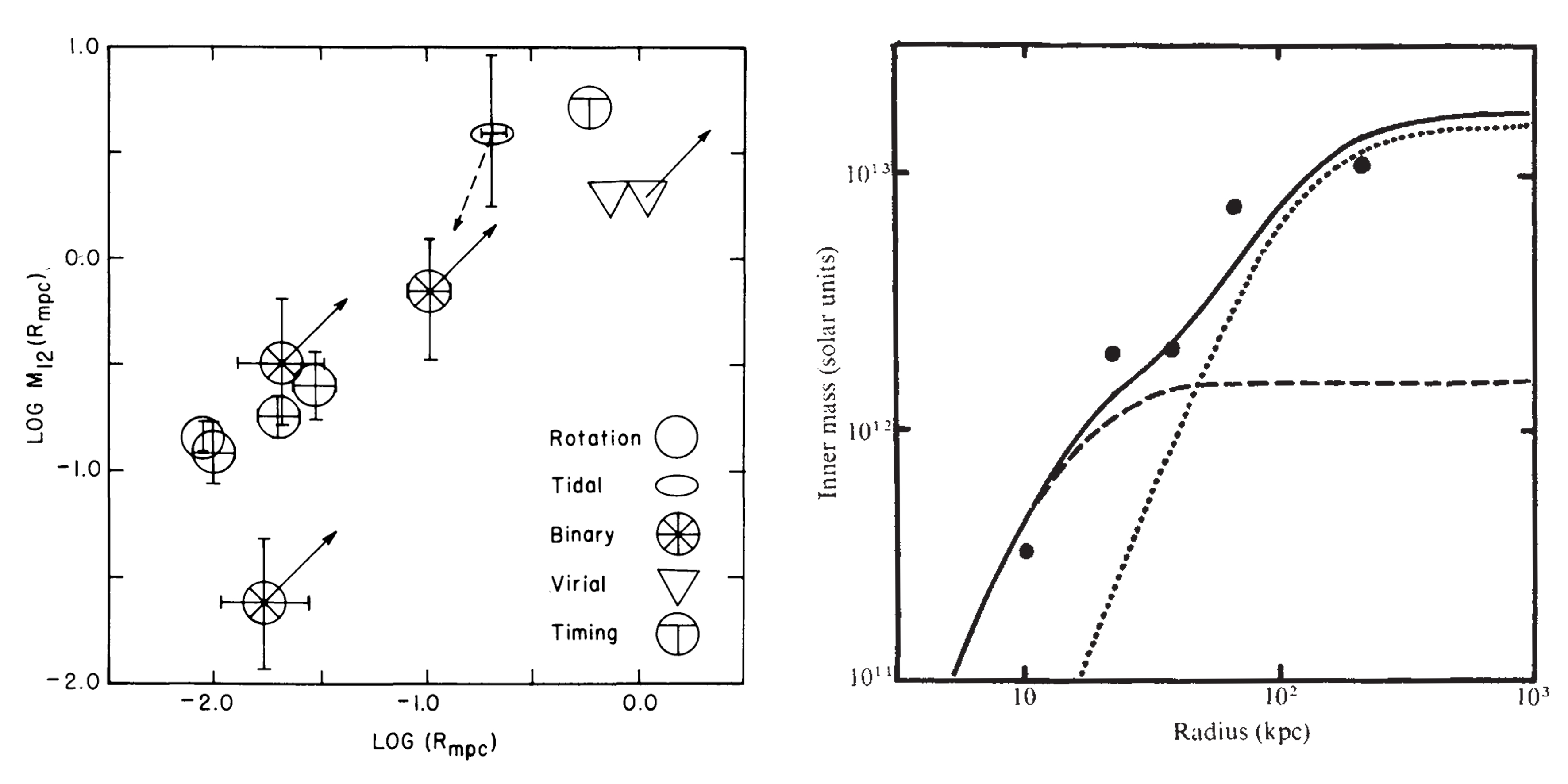}
    \caption{ \label{fig:EinastoOstriker}
Two diagrams from 1974 that plot the relation between the mass and the radius of galactic systems. \textbf{Left}: the mass of spiral galaxies as a function of radius by Ostriker, Peebles and Yahil (1974), as determined by various methods. Mass is in units of $10^{12} M_{\bigodot}$. \textbf{Right}: the relation between mass and radius of Einasto, Kaasik and Saar (1974). The dots represent the observed values obtained from pairs of galaxies, on the basis of data of Page (1970) and Karachentsev (1966). The dashed line represents the mass function of known stellar populations; the dotted line is the implied mass distribution of the `dark' corona; the solid line is the total mass distribution. Reproduced from ref. 15, AAS/IOP (left); and ref. 16, Macmillan Publishers Ltd (right).}
\end{figure}

Motivated by similar arguments, an Estonian group at Tartu Obervatory, consisting of Jaan Einasto, Ants Kaasik and Enn Saar, likewise concluded that the total mass density of matter in galaxies is 20 percent of the critical cosmological density.\cite{Einasto1974} For their influential paper (sent to \emph{Nature} a few weeks before Ostriker et al. would submit their work---both articles came out months later), the Estonians used rotation curve data of Roberts,\cite{Roberts1975} and masses of pairs of galaxies due to Thornton Page\cite{Page1970} and Igor Karachentsev\cite{Karachentsev1966}, among others. From these data and their own, Einasto and his co-workers constructed a diagram that plotted galaxy mass to radius similar to that of the Princeton group, which showed the value of the extra mass a dark corona surrounding a galaxy should have (see Fig. \ref{fig:EinastoOstriker}).

The Estonian group, just like its Princeton counterpart, was interdisciplinary in interest and background: astronomers and theoretical physicists joined efforts to study a problem that was now shared between galactic dynamics and cosmology. So, the authors of the 1974 papers were typical representatives of the new hybrid culture of physical cosmology; these were collaborations of differently trained scientists, working on novel energy scales and distances that were much larger than the familiar scales of their respective disciplines. The 1974 papers synthesized the two instances of curious galaxy behaviour into a single framework, thereby coalescing the problems into a single anomaly of missing mass. In fact, cosmology and its desire for a closed universe had turned the argument upside down: rather than that two autonomous problems had found a single solution, the discrepancies themselves started to function as \emph{evidence} for the existence of much wished for ``missing mass'', while different sets of data were put together as having a common origin.\cite{Janssen2002}

\begin{figure}[t]
\centering
	\includegraphics[width=0.8\textwidth]{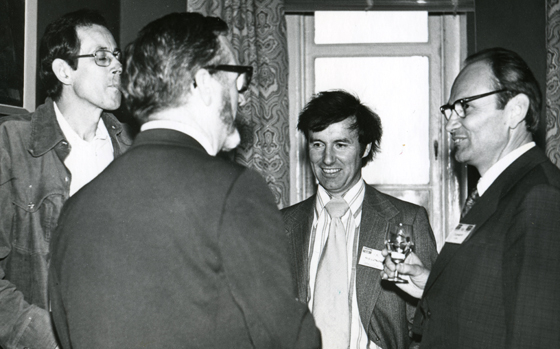}
\caption{Jaan Einasto and James Peebles together with George Abell and Malcom Longair at "The Large Scale Structure of Spacetime", the International Astronomical Union Symposium in Tallinn, Estonia/USSR (1977). Abell was known to support large mass-to-light ratios for clusters of galaxies already in the early 1960s,\cite{Abell1961} and Longair was among the early pioneers of physical cosmology.\cite{Longair1971} From left to right: Peebles, Abell, Longair and Einasto; source:  collection of J. Einasto.}
\end{figure}

\vspace{8pt}


\section*{Conclusion}

In the early 1970s, James Peebles recalls, ``a lot of things were not understood about masses of astronomical objects on the scales of galaxies and larger," (P.J.E. Peebles, oral history interview with J.G. de Swart, 13 November 2014) as cosmology offered a strong incentive to search for additional mass.
Additional mass was needed to close the universe, which in turn was desirable on (meta-)theoretical grounds. 
This new context and the new institutional expansion that accompanied it made the open problems of rotation curves and galaxy dynamics suddenly very prominent, even though both had been around for decades already. Now, the different problems were recognized as having a single origin.

Can the story of dark matter, as we have told it, inform us in current pertinent debates on the practice of cosmology, in particular regarding its methodology? Recently, authoritative cosmologists George Ellis and Joseph Silk have expressed strong concern about a perceived ``overclaiming'' of the significance of theory in modern cosmology;\cite{Ellis2014} as an antidote, they have proposed that Karl Popper's demarcation criterion of falsification should be reinstated:\cite{Popper1963} that science should consider methodological litmus tests of theories and their models to see whether they can be empirically falsified or not in order to decide whether they are deservedly included in our scientific discourse, and in the study of cosmology in particular.

First, one may ask how Popper's idea fared at the time that he argued for it. A prominent critic of Popper's philosophy was historian Thomas Kuhn, who pointed out that science actually does not usually progress via falsification tests; that, for instance,  on many occasions, theoretical innovation preceded observational reinterpretation, and that even at the moment of such reinterpretation, observations can look entirely equivocal, hence provide far from obvious `tests'---Kuhn's analysis of the Copernican revolution provided a forceful case in point.\cite{Kuhn1970} Historiography of science has moved on since Kuhn, but not in Popper's direction: as we have learned that the progress of science takes place along many dimensions (e.g. material instrumentation, theory, experiment, observation, and much else besides), it has become yet more clear that its history exhibits nuances that are not captured by Popperian prescriptions. In the same vein, the story of dark matter can not be reduced to a narrative in which `theories' or their `elegance' were the sole guide, nor indeed can it be framed as an exemplar of falsificationalism---its many interlocking components suggest a much richer texture that should forewarn one from constricting debate along the lines of `testability'.

Likewise, representations of the establishment of dark matter in terms of an `accumulation of evidence' miss an essential part of this history: they overlook the necessary \textit{conditions} that made this very accumulation possible; an accumulation, that at face value was substantially, even if partly, a reinterpretation of existing observations. Here, theory, along with an institutional shift and expansion in astronomy, played a substantive role: a role that reminds us that simply asking `what was the evidence' for missing matter, misses the point; we need to understand why certain observations were eventually conceived as `evidence' of anything in the first place.

Of course, the field of dark matter has dramatically evolved meanwhile, and much is still to be understood about how the problem travelled between different communities of scientists since 1974. The novel involvement of particle physics in the 1980s, for example, opened a whole new chapter that increased the significance and expanded the visibility of the problem. Clearly, further historical study may yet deepen our understanding of the actual practice and methods of physics, astronomy and cosmology---and assist us in navigating the current debates about the nature of dark matter.

\vspace{38pt}



\footnotesize{
 \bibliography{Bibliography}

\begin{thebibliography}{100}
\expandafter\ifx\csname url\endcsname\relax
  \def\url#1{\texttt{#1}}\fi
\expandafter\ifx\csname urlprefix\endcsname\relax\def\urlprefix{URL }\fi
\providecommand{\bibinfo}[2]{#2}
\providecommand{\eprint}[2][]{\url{#2}}

\bibitem{Bertone2010}
\bibinfo{author}{Bertone, G.}
\newblock \emph{\bibinfo{title}{{Particle Dark Matter: Observations, Models and
  Searches}}} (\bibinfo{publisher}{Cambridge University Press},
  \bibinfo{address}{Cambridge}, \bibinfo{year}{2010}).
\newblock

\bibitem{Trimble1987}
\bibinfo{author}{Trimble, V.}
\newblock \bibinfo{title}{{Existence and Nature of Dark Matter in the
  Universe}}.
\newblock \emph{\bibinfo{journal}{Annual Review of Astronomy and Astrophysics}}
  \textbf{\bibinfo{volume}{25}}, \bibinfo{pages}{425--472}
  (\bibinfo{year}{1987}).
\newblock 

\bibitem{Trimble2013}
\bibinfo{author}{Trimble, V.}
\newblock \bibinfo{title}{{History of Dark Matter in Galaxies}}.
\newblock In \bibinfo{editor}{Oswalt, T.~D.} \& \bibinfo{editor}{Gilmore, G.}
  (eds.) \emph{\bibinfo{booktitle}{Planets, Stars and Stellar Systems Vol. 5}},
  vol.~\bibinfo{volume}{5}, \bibinfo{pages}{1091} (\bibinfo{publisher}{Springer
  Netherlands}, \bibinfo{address}{Dordrecht}, \bibinfo{year}{2013}).
\newblock 

\bibitem{Sanders2010}
\bibinfo{author}{Sanders, R.~H.}
\newblock \emph{\bibinfo{title}{{The Dark Matter Problem: A Historical
  Perspective}}} (\bibinfo{publisher}{Cambridge University Press},
  \bibinfo{address}{Cambridge}, \bibinfo{year}{2010}).
\newblock

\bibitem{Einasto2014}
\bibinfo{author}{Einasto, J.}
\newblock \emph{\bibinfo{title}{{Dark Matter and Cosmic Web Story}}}
  (\bibinfo{publisher}{World Scientific Publishing},
  \bibinfo{address}{Singapore}, \bibinfo{year}{2014}).
\newblock

\bibitem{Bertone2016}
\bibinfo{author}{Bertone, G.} \& \bibinfo{author}{Hooper, D.}
\newblock \bibinfo{title}{{A History of Dark Matter}}.
\newblock \emph{\bibinfo{journal}{Submitted to: Rev. Mod. Phys.}}
  (\bibinfo{year}{2016}).
\newblock 
\newblock \eprint{1605.04909}.

\bibitem{Kapteyn1922}
\bibinfo{author}{Kapteyn, J.~C.}
\newblock \bibinfo{title}{{First Attempt at a Theory of the Arrangement and
  Motion of the Sidereal System}}.
\newblock \emph{\bibinfo{journal}{The Astrophysical Journal}}
  \textbf{\bibinfo{volume}{55}}, \bibinfo{pages}{302} (\bibinfo{year}{1922}).
\newblock 

\bibitem{Jeans1922}
\bibinfo{author}{Jeans, J.~H.}
\newblock \bibinfo{title}{{The Motions of Stars in a Kapteyn-Universe}}.
\newblock \emph{\bibinfo{journal}{Monthly Notices of the Royal Astronomical
  Society}} \textbf{\bibinfo{volume}{82}}, \bibinfo{pages}{122--132}
  (\bibinfo{year}{1922}).
\newblock

\bibitem{Oort1927}
\bibinfo{author}{Oort, J.~H.}
\newblock \bibinfo{title}{{Observational evidence confirming Lindblad's
  hypothesis of a rotation of the galactic system}}.
\newblock \emph{\bibinfo{journal}{Bulletin of the Astronomical Institutes of
  the Netherlands}} \textbf{\bibinfo{volume}{3}} (\bibinfo{year}{1927}).
\newblock 

\bibitem{Zwicky1933}
\bibinfo{author}{Zwicky, F.}
\newblock \bibinfo{title}{{Die Rotverschiebung von extragalaktischen Nebeln}}.
\newblock \emph{\bibinfo{journal}{Helvetica Physica Acta}}
  \textbf{\bibinfo{volume}{6}}, \bibinfo{pages}{110--127}
  (\bibinfo{year}{1933}).
\newblock 
\newblock \eprint{arXiv:1011.1669v3}.

\bibitem{Smith1936a}
\bibinfo{author}{Smith, S.}
\newblock \bibinfo{title}{{The mass of the Virgo cluster.}}
\newblock \emph{\bibinfo{journal}{Contributions from the Mount Wilson
  Observatory / Carnegie Institution of Washington}}
  \textbf{\bibinfo{volume}{532}} (\bibinfo{year}{1936}).
\newblock 

\bibitem{Holmberg1937}
\bibinfo{author}{Holmberg, E.}
\newblock \bibinfo{title}{{A Study of Double and Multiple Galaxies Together
  with Inquiries into some General Metagalactic Problems}}.
\newblock \emph{\bibinfo{journal}{Annals of the Observatory of Lund}}
  \textbf{\bibinfo{volume}{6}} (\bibinfo{year}{1937}).
\newblock 

\bibitem{Faber1979}
\bibinfo{author}{Faber, S.~M.} \& \bibinfo{author}{Gallagher, J.~S.}
\newblock \bibinfo{title}{{Masses and Mass-To-Light Ratios of Galaxies}}.
\newblock \emph{\bibinfo{journal}{Annual Review of Astronomy and Astrophysics}}
  \textbf{\bibinfo{volume}{17}}, \bibinfo{pages}{135--187}
  (\bibinfo{year}{1979}).
\newblock

\bibitem{Riess2015}
\bibinfo{author}{Riess, A.}
\newblock \bibinfo{title}{{Dark Matter}}.
\newblock In \emph{\bibinfo{booktitle}{Encyclop{\ae}dia Britannica.}}
  (\bibinfo{publisher}{Retrieved from
  http://www.britannica.com/science/dark-matter}, \bibinfo{year}{2015}).
\newblock 

\bibitem{Ostriker1974}
\bibinfo{author}{Ostriker, J.~P.}, \bibinfo{author}{Peebles, P. J.~E.} \&
  \bibinfo{author}{Yahil, A.}
\newblock \bibinfo{title}{{The size and mass of galaxies, and the mass of the
  universe}}.
\newblock \emph{\bibinfo{journal}{Astrophysical Journal}}
  \textbf{\bibinfo{volume}{193}}, \bibinfo{pages}{L1} (\bibinfo{year}{1974}).
\newblock

\bibitem{Einasto1974}
\bibinfo{author}{Einasto, J.}, \bibinfo{author}{Kaasik, A.} \&
  \bibinfo{author}{Saar, E.}
\newblock \bibinfo{title}{{Dynamic evidence on massive coronas of galaxies}}.
\newblock \emph{\bibinfo{journal}{Nature}} \textbf{\bibinfo{volume}{250}},
  \bibinfo{pages}{309--310} (\bibinfo{year}{1974}).
\newblock

\bibitem{Zwicky1956a}
\bibinfo{author}{Zwicky, F.}
\newblock \bibinfo{title}{{Statistics of Clusters of Galaxies. 
  }}
\newblock \emph{\bibinfo{journal}{Proceedings of the Third Berkeley Symposium
  on Mathematical Statistics and Probability}} \textbf{\bibinfo{volume}{3}}
  (\bibinfo{year}{1956}).
\newblock 

\bibitem{Shane1956}
\bibinfo{author}{Shane, C.~D.} \& \bibinfo{author}{Wirtanen, C.~A.}
\newblock \bibinfo{title}{{The distribution of extragalactic nebulae}}.
\newblock \emph{\bibinfo{journal}{Astronomical Journal}}
  \textbf{\bibinfo{volume}{59}}, \bibinfo{pages}{285} (\bibinfo{year}{1954}).
\newblock

\bibitem{Abell1957}
\bibinfo{author}{Abell, G.~O.}
\newblock \emph{\bibinfo{title}{{The Distribution of Rich Clusters of
  Galaxies.}}}
\newblock Ph.D. thesis, \bibinfo{school}{California Institute of Technology}
  (\bibinfo{year}{1958}).
\newblock 


\bibitem{Abell1959}
\bibinfo{author}{Abell, G.~O.}
\newblock \bibinfo{title}{{The National Geographic Society-Palomar
  Observatory Sky Survey}}.
\newblock \emph{\bibinfo{journal}{Astronomical Society of the Pacific Leaflets}}
  \textbf{\bibinfo{volume}{8}}, \bibinfo{pages}{121} (\bibinfo{year}{1954}).
\newblock


\bibitem{Kahn1959}
\bibinfo{author}{Kahn, F.~D.} \& \bibinfo{author}{Woltjer, L.}
\newblock \bibinfo{title}{{Intergalactic Matter and the Galaxy.}}
\newblock \emph{\bibinfo{journal}{The Astrophysical Journal}}
  \textbf{\bibinfo{volume}{130}}, \bibinfo{pages}{705} (\bibinfo{year}{1959}).
\newblock 

\bibitem{Page1959}
\bibinfo{author}{Page, T.}
\newblock \bibinfo{title}{{Masses of the double galaxies.}}
\newblock \emph{\bibinfo{journal}{The Astronomical Journal}}
  \textbf{\bibinfo{volume}{64}}, \bibinfo{pages}{53} (\bibinfo{year}{1959}).
\newblock

\bibitem{Ambartsumian1958a}
\bibinfo{author}{Ambartsumian, V.~A.}
\newblock \bibinfo{title}{{On the Evolution of Galaxies}}.
\newblock In \bibinfo{editor}{Stoops, R.} (ed.) \emph{\bibinfo{booktitle}{La
  structure et l'evolution de l'universe}}, \bibinfo{pages}{241--279}
  (\bibinfo{publisher}{Onzi{\`{e}}me Conseil de Physique},
  \bibinfo{address}{Bruxelles}, \bibinfo{year}{1958}).

\bibitem{Burbidge1959}
\bibinfo{author}{Burbidge, G.~R.} \& \bibinfo{author}{Burbidge, E.~M.}
\newblock \bibinfo{title}{{The Hercules Clusters of Nebulae.}}
\newblock \emph{\bibinfo{journal}{The Astrophysical Journal}}
  \textbf{\bibinfo{volume}{130}}, \bibinfo{pages}{629} (\bibinfo{year}{1959}).
\newblock 

\bibitem{VandenBergh1961}
\bibinfo{author}{van~den Bergh, S.}
\newblock \bibinfo{title}{{The Stability of Clusters of Galaxies}}.
\newblock \emph{\bibinfo{journal}{The Astronomical Journal}}
  \textbf{\bibinfo{volume}{66}}, \bibinfo{pages}{566} (\bibinfo{year}{1961}).
\newblock 

\bibitem{Neyman1961a}
\bibinfo{author}{Neyman, J.}, \bibinfo{author}{Page, T.} \&
  \bibinfo{author}{Scott, E.}
\newblock \bibinfo{title}{{Conference on the Instability of Systems of Galaxies
  (Santa Barbara, California, August 10-12, 1961): Foreword}}.
\newblock \emph{\bibinfo{journal}{The Astronomical Journal}}
  \textbf{\bibinfo{volume}{66}}, \bibinfo{pages}{533} (\bibinfo{year}{1961}).

\bibitem{Burbidge1962}
\bibinfo{author}{Burbidge, G.~R.}
\newblock \bibinfo{title}{{Multiple Systems, Clusters, Radiogalaxies:
  Summary}}.
\newblock In \bibinfo{editor}{McVittie, G.} (ed.)
  \emph{\bibinfo{booktitle}{Problems of Extra-Galactic Research, Proceedings
  from IAU Symposium no. 15}}, \bibinfo{pages}{258--265}
  (\bibinfo{publisher}{Macmillan}, \bibinfo{address}{New York},
  \bibinfo{year}{1962}).

\bibitem{Woolf1967}
\bibinfo{author}{Woolf, N.~J.}
\newblock \bibinfo{title}{{On the Stabilization of Clusters of Galaxies by
  Ionized Gas}}.
\newblock \emph{\bibinfo{journal}{The Astrophysical Journal}}
  \textbf{\bibinfo{volume}{148}}, \bibinfo{pages}{287} (\bibinfo{year}{1967}).
\newblock 

\bibitem{Reddish1968}
\bibinfo{author}{Reddish, V.~C.}
\newblock \bibinfo{title}{{The Evolution of Galaxies.}}
\newblock \emph{\bibinfo{journal}{Quarterly Journal of the Royal Astronomical Society}}
  \textbf{\bibinfo{volume}{9}}, \bibinfo{pages}{416} (\bibinfo{year}{1968}).
\newblock 
\newblock 


\bibitem{Finzi1963}
\bibinfo{author}{Finzi, A.}
\newblock \bibinfo{title}{{On the validity of Newton's law at a long
  distance}}.
\newblock \emph{\bibinfo{journal}{Monthly Notices of the Royal Astronomical
  Society}} \textbf{\bibinfo{volume}{127}} (\bibinfo{year}{1963}).
\newblock 

\bibitem{Forman1970}
\bibinfo{author}{Forman, W.~R.}
\newblock \bibinfo{title}{{A Reduction of the Mass Deficit in Clusters of
  Galaxies by Means of a Negative Cosmological Constant}}.
\newblock \emph{\bibinfo{journal}{The Astrophysical Journal}}
  \textbf{\bibinfo{volume}{159}}, \bibinfo{pages}{719--722}
  (\bibinfo{year}{1970}).
\newblock 

\bibitem{Jackson1970}
\bibinfo{author}{Jackson, J.~C.}
\newblock \bibinfo{title}{{The dynamics of clusters of galaxies in universes
  with non-zero cosmological constant, and the virial theorem mass
  discrepancy}}.
\newblock \emph{\bibinfo{journal}{Monthly Notices of the Royal Astronomical
  Society}} \textbf{\bibinfo{volume}{148}}, \bibinfo{pages}{249--260}
  (\bibinfo{year}{1970}).
\newblock 

\bibitem{Field1971}
\bibinfo{author}{Field, G.~B.} \& \bibinfo{author}{Saslaw, W.~C.}
\newblock \bibinfo{title}{{Groups of Galaxies: Hidden Mass or Quick
  Disintegration?}}
\newblock \emph{\bibinfo{journal}{The Astrophysical Journal}}
  \textbf{\bibinfo{volume}{170}}, \bibinfo{pages}{199} (\bibinfo{year}{1971}).
\newblock 

\bibitem{Hawking1971}
\bibinfo{author}{Hawking, S.}
\newblock \bibinfo{title}{{Gravitationally Collapsed Objects of Very Low
  Mass}}.
\newblock \emph{\bibinfo{journal}{Monthly Notices of the Royal Astronomical
  Society}} \textbf{\bibinfo{volume}{152}}, \bibinfo{pages}{75--78}
  (\bibinfo{year}{1971}).
\newblock 

\bibitem{Gott1973}
\bibinfo{author}{Gott, I., J.~Richard}, \bibinfo{author}{Wrixon, G.~T.} \&
  \bibinfo{author}{Wannier, P.}
\newblock \bibinfo{title}{{A Study of Three Groups of Galaxies: Plausible
  Explanation of the Virial Mass Discrepancy}}.
\newblock \emph{\bibinfo{journal}{The Astrophysical Journal}}
  \textbf{\bibinfo{volume}{186}}, \bibinfo{pages}{777} (\bibinfo{year}{1973}).
\newblock 

\bibitem{Cowsik1972}
\bibinfo{author}{Cowsik, R.} \& \bibinfo{author}{McClelland, J.}
\newblock \bibinfo{title}{{An upper limit on the neutrino rest mass}}.
\newblock \emph{\bibinfo{journal}{Physical Review Letters}}
  \textbf{\bibinfo{volume}{29}}, \bibinfo{pages}{669--670}
  (\bibinfo{year}{1973}).
\newblock 

\bibitem{Abell1975}
\bibinfo{author}{Abell, G.~O.}
\newblock \bibinfo{title}{{Clusters of Galaxies}}.
\newblock In \bibinfo{editor}{Sandage, A.}, \bibinfo{editor}{Sandage, M.} \&
  \bibinfo{editor}{Kristian, J.} (eds.) \emph{\bibinfo{booktitle}{Stars and
  Stellar Systems: Galaxies and the Universe.}}, vol.~\bibinfo{volume}{9},
  chap.~\bibinfo{chapter}{15}, \bibinfo{pages}{601--646}
  (\bibinfo{publisher}{The University of Chicago Press}, \bibinfo{year}{1975}).
\newblock 

\bibitem{Karachentsev1966}
\bibinfo{author}{Karachentsev, I.~D.}
\newblock \bibinfo{title}{{The virial mass-luminosity ratio and the instability
  of different galactic systems}}.
\newblock \emph{\bibinfo{journal}{Astrophysics}} \textbf{\bibinfo{volume}{2}},
  \bibinfo{pages}{39--49} (\bibinfo{year}{1966}).
\newblock 

\bibitem{Burbidge1969}
\bibinfo{author}{Burbidge, G.~R.} \& \bibinfo{author}{Sargent, W. L.~W.}
\newblock \bibinfo{title}{{The Case of the Missing Mass}}.
\newblock \emph{\bibinfo{journal}{Comments on Astrophysics and Space Physics}}
  \textbf{\bibinfo{volume}{1}} (\bibinfo{year}{1969}).
\newblock 

\bibitem{DeVaucouleurs1975}
\bibinfo{author}{de~Vaucouleurs, G.}
\newblock \bibinfo{title}{{Nearby Groups of Galaxies}}.
\newblock In \bibinfo{editor}{Sandage, A.}, \bibinfo{editor}{Sandage, M.} \&
  \bibinfo{editor}{Kristian, J.} (eds.) \emph{\bibinfo{booktitle}{Stars and
  Stellar Systems: Galaxies and the Universe.}}, vol.~\bibinfo{volume}{9},
  chap.~\bibinfo{chapter}{14} (\bibinfo{publisher}{The University of Chicago
  Press}, \bibinfo{year}{1975}).
\newblock 

\bibitem{Lindblad1927}
\bibinfo{author}{Lindblad, B.}
\newblock \bibinfo{title}{{On the State of Motion in the Galactic System}}.
\newblock \emph{\bibinfo{journal}{Monthly Notices of the Royal Astronomical
  Society}} \textbf{\bibinfo{volume}{87}}, \bibinfo{pages}{553--564}
  (\bibinfo{year}{1927}).
\newblock 

\bibitem{Babcock1939}
\bibinfo{author}{Babcock, H.~W.}
\newblock \bibinfo{title}{{The Rotation of the Andromeda Nebula}}.
\newblock \emph{\bibinfo{journal}{Lick Observatory Bulletin}}
  \textbf{\bibinfo{volume}{19}}, \bibinfo{pages}{41--51}
  (\bibinfo{year}{1939}).
\newblock 

\bibitem{Mayall1942}
\bibinfo{author}{Mayall, N.} \& \bibinfo{author}{Aller, L.}
\newblock \bibinfo{title}{{The Rotation of the Spiral Nebula Messier 33.}}
\newblock \emph{\bibinfo{journal}{Astrophysical Journal}}
  \textbf{\bibinfo{volume}{95}}, \bibinfo{pages}{5} (\bibinfo{year}{1942}).
\newblock 

\bibitem{Ewen1951}
\bibinfo{author}{Ewen, H.~I.} \& \bibinfo{author}{Purcell, E.~M.}
\newblock \bibinfo{title}{{Observation of a line in the galactic radio
  spectrum}}.
\newblock \emph{\bibinfo{journal}{Nature}} \textbf{\bibinfo{volume}{168}},
  \bibinfo{pages}{356} (\bibinfo{year}{1951}).
\newblock 

\bibitem{Muller1951}
\bibinfo{author}{Muller, C.~A.} \& \bibinfo{author}{Oort, J.~H.}
\newblock \bibinfo{title}{{The Interstellar Hydrogen Line at 1,420 Mc./sec, and
  an Estimate of Galactic Rotation}}.
\newblock \emph{\bibinfo{journal}{Nature}} \textbf{\bibinfo{volume}{168}},
  \bibinfo{pages}{357--358} (\bibinfo{year}{1951}).
\newblock 

\bibitem{Pawsey1951}
\bibinfo{author}{Pawsey, J.~L.}
\newblock \bibinfo{title}{{Observation of a Line in the Galactic Radio
  Spectrum: The Interstellar Hydrogen Line at 1,420 Mc./sec., and an Estimate
  of Galactic Rotation}}.
\newblock \emph{\bibinfo{journal}{Nature}} \textbf{\bibinfo{volume}{168}},
  \bibinfo{pages}{358--358} (\bibinfo{year}{1951}).
\newblock 

\bibitem{Freeman1970}
\bibinfo{author}{Freeman, K.~C.}
\newblock \bibinfo{title}{{On the Disks of Spiral and S0 Galaxies}}.
\newblock \emph{\bibinfo{journal}{Astrophysical Journal}}
  \textbf{\bibinfo{volume}{160}}, \bibinfo{pages}{811} (\bibinfo{year}{1970}).
\newblock

\bibitem{Rubin1970}
\bibinfo{author}{Rubin, V.~C.} \& \bibinfo{author}{Ford, J., W.~Kent}.
\newblock \bibinfo{title}{{Rotation of the Andromeda Nebula from a
  Spectroscopic Survey of Emission Regions}}.
\newblock \emph{\bibinfo{journal}{The Astrophysical Journal}}
  \textbf{\bibinfo{volume}{159}}, \bibinfo{pages}{379} (\bibinfo{year}{1970}).
\newblock 

\bibitem{Roberts1973}
\bibinfo{author}{Roberts, M.~S.} \& \bibinfo{author}{Rots, A.~H.}
\newblock \bibinfo{title}{{Comparison of Rotation Curves of Different Galaxy
  Types}}.
\newblock \emph{\bibinfo{journal}{Astronomy and Astrophysics}}
  \textbf{\bibinfo{volume}{26}}, \bibinfo{pages}{483--485}
  (\bibinfo{year}{1973}).
\newblock

\bibitem{Shostak1972}
\bibinfo{author}{Shostak, G.~S.}
\newblock \emph{\bibinfo{title}{{Aperture Synthesis Observations of Neutral
  Hydrogen in Three Galaxies}}}.
\newblock Ph.D. thesis (\bibinfo{year}{1972}).
\newblock 

\bibitem{Rogstad1972}
\bibinfo{author}{Rogstad, D.~H.} \& \bibinfo{author}{Shostak, G.~S.}
\newblock \bibinfo{title}{{Gross Properties of Five Scd Galaxies as Determined
  from 21-cm Observations}}.
\newblock \emph{\bibinfo{journal}{The Astrophysical Journal}}
  \textbf{\bibinfo{volume}{176}}, \bibinfo{pages}{315} (\bibinfo{year}{1972}).
\newblock 

\bibitem{Rogstad1973}
\bibinfo{author}{Rogstad, D.~H.}, \bibinfo{author}{Shostak, G.~S.} \&
  \bibinfo{author}{Rots, A.~H.}
\newblock \bibinfo{title}{{Aperture synthesis study of neutral hydrogen in the
  galaxies NGC 6946 and IC 342.}}
\newblock \emph{\bibinfo{journal}{Astronomy and Astrophysics}}
  \textbf{\bibinfo{volume}{22}}, \bibinfo{pages}{111--119}
  (\bibinfo{year}{1973}).
\newblock

\bibitem{Brandt1960}
\bibinfo{author}{Brandt, J.~C.}
\newblock \bibinfo{title}{{On the Distribution Of Mass in Galaxies. I. The
  Large-Scale Structure of Ordinary Spirals with Applications to M 31.}}
\newblock \emph{\bibinfo{journal}{The Astrophysical Journal}}
  \textbf{\bibinfo{volume}{131}}, \bibinfo{pages}{293} (\bibinfo{year}{1960}).
\newblock

\bibitem{Toomre1963}
\bibinfo{author}{Toomre, A.}
\newblock \bibinfo{title}{{On the Distribution of Matter Within Highly
  Flattened Galaxies.}}
\newblock \emph{\bibinfo{journal}{The Astrophysical Journal}}
  \textbf{\bibinfo{volume}{138}}, \bibinfo{pages}{385} (\bibinfo{year}{1963}).
\newblock 

\bibitem{Huchtmeier1975}
\bibinfo{author}{Huchtmeier, W.}
\newblock \bibinfo{title}{{Rotation-curves of galaxies from 21 cm-line
  observations}}.
\newblock \emph{\bibinfo{journal}{Astronomy and Astrophysics}}
  \textbf{\bibinfo{volume}{45}}, \bibinfo{pages}{259--268}
  (\bibinfo{year}{1975}).
\newblock

\bibitem{Emerson1973}
\bibinfo{author}{Emerson, D.~T.} \& \bibinfo{author}{Baldwin, J.~E.}
\newblock \bibinfo{title}{{The Rotation Curve and Mass Distribution in M31}}.
\newblock \emph{\bibinfo{journal}{Monthly Notices of the Royal Astronomical
  Society}} \textbf{\bibinfo{volume}{165}}, \bibinfo{pages}{9P--13P}
  (\bibinfo{year}{1973}).
\newblock 

\bibitem{Baldwin1975}
\bibinfo{author}{Baldwin, J.~E.}
\newblock \bibinfo{title}{{M/L Ratios in Galactic Disks}}.
\newblock In \bibinfo{editor}{Hayli, A.} (ed.)
  \emph{\bibinfo{booktitle}{Dynamics of Stellar Systems: Proceedings from IAU
  Symposium no. 69 held in Besancon}}, \bibinfo{pages}{341--348}
  (\bibinfo{publisher}{D. Reidel}, \bibinfo{address}{Dordrecht},
  \bibinfo{year}{1975}).
\newblock 

\bibitem{Roberts1975}
\bibinfo{author}{Roberts, M.~S.}
\newblock \bibinfo{title}{{Radio Observations of Neutral Hydrogen in
  Galaxies}}.
\newblock In \bibinfo{editor}{Sandage, A.}, \bibinfo{editor}{Sandage, M.} \&
  \bibinfo{editor}{Kristian, J.} (eds.) \emph{\bibinfo{booktitle}{Galaxies and
  the Universe.}}, vol.~\bibinfo{volume}{9}, chap.~\bibinfo{chapter}{9},
  \bibinfo{pages}{309--358} (\bibinfo{publisher}{The University of Chicago
  Press}, \bibinfo{address}{Chicago}, \bibinfo{year}{1975}).
\newblock 

\bibitem{Schwarzschild1954}
\bibinfo{author}{Schwarzschild, M.}
\newblock \bibinfo{title}{{Mass distribution and mass-luminosity ratio in
  galaxies}}.
\newblock \emph{\bibinfo{journal}{The Astronomical Journal}}
  \textbf{\bibinfo{volume}{59}}, \bibinfo{pages}{273} (\bibinfo{year}{1954}).
\newblock

\bibitem{Smith2008a}
\bibinfo{author}{Smith, R.~W.}
\newblock \bibinfo{title}{{Beyond the Galaxy: The Development of Extragalactic
  Astronomy 1885-1965, Part 1}}.
\newblock \emph{\bibinfo{journal}{Journal for the History of Astronomy}}
  \textbf{\bibinfo{volume}{39}}, \bibinfo{pages}{91--119}
  (\bibinfo{year}{2008}).
\newblock 

\bibitem{Smith2009}
\bibinfo{author}{Smith, R.~W.}
\newblock \bibinfo{title}{{Beyond the Galaxy: The Development of Extragalactic
  Astronomy 1885-1965, Part 2}}.
\newblock \emph{\bibinfo{journal}{Journal for the History of Astronomy}}
  \textbf{\bibinfo{volume}{40}}, \bibinfo{pages}{71--107}
  (\bibinfo{year}{2009}).
\newblock 

\bibitem{Roberts1965}
\bibinfo{author}{Roberts, M.~S.}
\newblock \bibinfo{title}{{Recent Discoveries in radio astronomy}}.
\newblock \emph{\bibinfo{journal}{Physics Today}}
  \textbf{\bibinfo{volume}{18}}, \bibinfo{pages}{28--36}
  (\bibinfo{year}{1965}).
\newblock 

\bibitem{Schmidt1963}
\bibinfo{author}{Schmidt, M.}
\newblock \bibinfo{title}{{3C 273 : A Star-Like Object with Large Red-Shift}}.
\newblock \emph{\bibinfo{journal}{Nature}} \textbf{\bibinfo{volume}{197}},
  \bibinfo{pages}{1040} (\bibinfo{year}{1963}).
\newblock 

\bibitem{Kragh2006}
\bibinfo{author}{Kragh, H.~S.}
\newblock \emph{\bibinfo{title}{{Conceptions of cosmos: from myths to the
  accelerating universe: a history of cosmology}}} (\bibinfo{publisher}{Oxford
  University Press}, \bibinfo{address}{Oxford}, \bibinfo{year}{2006}).
\newblock

\bibitem{Eisenstaedt2006}
\bibinfo{author}{Eisenstaedt, J.}
\newblock \emph{\bibinfo{title}{{The Curious History of Relativity: How
  Einstein's Theory of Gravity Was Lost and Found Again}}}
  (\bibinfo{publisher}{Princeton University Press},
  \bibinfo{address}{Princeton, NJ}, \bibinfo{year}{2006}).

\bibitem{Dongen2010}
\bibinfo{author}{van Dongen, J.}
\newblock \emph{\bibinfo{title}{{Einstein's Unification}}}
  (\bibinfo{publisher}{Cambridge University Press},
  \bibinfo{address}{Cambridge}, \bibinfo{year}{2010}).
\newblock

\bibitem{Gold1965}
\bibinfo{author}{Gold, T.}
\newblock \bibinfo{title}{{After-Dinner Speech}}.
\newblock In \bibinfo{editor}{Robinson, I.}, \bibinfo{editor}{Schild, A.} \&
  \bibinfo{editor}{Schucking, E.~L.} (eds.)
  \emph{\bibinfo{booktitle}{Quasi-Stellar Sources and Gravitational Collapse,
  Proceedings of the 1st Texas Symposium on Relativistic Astrophysics}},
  \bibinfo{pages}{470} (\bibinfo{publisher}{University Of Chicago Press},
  \bibinfo{address}{Chicago}, \bibinfo{year}{1965}).
\newblock 

\bibitem{Blum2015}
\bibinfo{author}{Blum, A.}, \bibinfo{author}{Lalli, R.} \&
  \bibinfo{author}{Renn, J.}
\newblock \bibinfo{title}{{The Reinvention of General Relativity : A
  Historiographical Framework for Assessing One Hundred Years of Curved
  Space-time}}.
\newblock \emph{\bibinfo{journal}{Isis}} \textbf{\bibinfo{volume}{106}},
  \bibinfo{pages}{598--620} (\bibinfo{year}{2015}).
\newblock 

\bibitem{Sciama1971a}
\bibinfo{author}{Sciama, D.~W.}
\newblock \emph{\bibinfo{title}{{Modern Cosmology}}}
  (\bibinfo{publisher}{Cambridge: University Press},
  \bibinfo{address}{Cambridge}, \bibinfo{year}{1971}), \bibinfo{edition}{1}
  edn.
\newblock 

\bibitem{Page1964}
\bibinfo{author}{Dressler, A.}
\newblock \bibinfo{title}{{The Evolution of Galaxies in Clusters}}.
\newblock \emph{\bibinfo{journal}{Annual Review of Astronomy and Astrophysics}}
  \textbf{\bibinfo{volume}{22}}, \bibinfo{pages}{185--222}
  (\bibinfo{year}{1984}).
\newblock 

\bibitem{Tinsley1968a}
\bibinfo{author}{Tinsley, B.~M.}
\newblock \bibinfo{title}{{Evolution of the Stars and Gas in Galaxies}}.
\newblock \emph{\bibinfo{journal}{The Astrophysical Journal}}
  \textbf{\bibinfo{volume}{151}}, \bibinfo{pages}{547} (\bibinfo{year}{1968}).
\newblock 

\bibitem{Hewish1968}
\bibinfo{author}{Hewish, A.}, \bibinfo{author}{Bell, S.~J.},
  \bibinfo{author}{Pilkington, J. D.~H.}, \bibinfo{author}{Scott, P.~F.} \&
  \bibinfo{author}{Collins, R.~a.}
\newblock \bibinfo{title}{{Observation of a Rapidly Pulsating Radio Source}}.
\newblock \emph{\bibinfo{journal}{Nature}} \textbf{\bibinfo{volume}{224}},
  \bibinfo{pages}{472--472} (\bibinfo{year}{1968}).
\newblock 

\bibitem{NationalResearchCouncil1972}
\bibinfo{author}{{National Research Council}}.
\newblock \emph{\bibinfo{title}{{Astronomy and Astrophysics for the 1970's:
  Volume 1: Report of the Astronomy Survey Committee}}},
  vol.~\bibinfo{volume}{1} (\bibinfo{publisher}{National Academies Press},
  \bibinfo{address}{Washington, DC}, \bibinfo{year}{1972}).
\newblock

\bibitem{NationalResearchCouncil1964}
\bibinfo{author}{{National Research Council}}.
\newblock \emph{\bibinfo{title}{{Ground-based Astronomy: A Ten-year Program}}}
  (\bibinfo{publisher}{National Academies Press}, \bibinfo{address}{Washington,
  D.C.}, \bibinfo{year}{1964}).
\newblock

\bibitem{NationalResearchCouncil1973}
\bibinfo{author}{{National Research Council}}.
\newblock \emph{\bibinfo{title}{{Astronomy and Astrophysics for the 1970's:
  Volume 2: Report of the Panels}}} (\bibinfo{publisher}{National Academies
  Press}, \bibinfo{address}{Washington, D.C.}, \bibinfo{year}{1973}).
\newblock 

\bibitem{Kaiser2002}
\bibinfo{author}{Kaiser, D.}
\newblock \bibinfo{title}{{Cold War requisitions, scientific manpower, and the
  production of American physicists after World War II}}.
\newblock \emph{\bibinfo{journal}{Historical Studies in the Physical and
  Biological Sciences}} \textbf{\bibinfo{volume}{33}},
  \bibinfo{pages}{131--159} (\bibinfo{year}{2002}).
\newblock 

\bibitem{Kaiser2006}
\bibinfo{author}{Kaiser, D.}
\newblock \bibinfo{title}{{Whose Mass is it Anyway? Particle Cosmology and the
  Objects of Theory}}.
\newblock \emph{\bibinfo{journal}{Social Studies of Science}}
  \textbf{\bibinfo{volume}{36}}, \bibinfo{pages}{533--564}
  (\bibinfo{year}{2006}).
\newblock

\bibitem{NationalResearchCouncil1983}
\bibinfo{author}{{National Research Council}}.
\newblock \emph{\bibinfo{title}{{Astronomy and Astrophysics for the 1980's,
  Volume 2: Reports of the Panels}}} (\bibinfo{publisher}{National Academies
  Press}, \bibinfo{address}{Washington, D.C.}, \bibinfo{year}{1983}).

\bibitem{Peebles1971}
\bibinfo{author}{Peebles, P. J.~E.}
\newblock \emph{\bibinfo{title}{{Physical Cosmology}}}
  (\bibinfo{publisher}{Princeton University Press},
  \bibinfo{address}{Princeton, NJ}, \bibinfo{year}{1971}).
\newblock 

\bibitem{Weinberg1972}
\bibinfo{author}{Weinberg, S.}
\newblock \emph{\bibinfo{title}{{Gravitation and Cosmology: Principles and
  Applications of the General Theory of Relativity}}},
  vol.~\bibinfo{volume}{41} (\bibinfo{publisher}{John Wiley {\&} Sons, Inc.},
  \bibinfo{year}{1972}).
\newblock 

\bibitem{Misner1973}
\bibinfo{author}{Misner, C.~W.}, \bibinfo{author}{Thorne, K.~S.} \&
  \bibinfo{author}{Wheeler, J.~A.}
\newblock \emph{\bibinfo{title}{{Gravitation}}} (\bibinfo{publisher}{W.H.
  Freeman and Co.}, \bibinfo{address}{San Francisco, CA},
  \bibinfo{year}{1973}).
\newblock 

\bibitem{Hawking1973}
\bibinfo{author}{Hawking, S.~W.} \& \bibinfo{author}{Ellis, G. F.~R.}
\newblock \emph{\bibinfo{title}{{The large-scale structure of space-time.}}}
  (\bibinfo{publisher}{Cambridge University Press},
  \bibinfo{address}{Cambridge}, \bibinfo{year}{1973}).
\newblock 

\bibitem{Sandage1970}
\bibinfo{author}{Sandage, A.~R.}
\newblock \bibinfo{title}{{Cosmology: A search for two numbers}}.
\newblock \emph{\bibinfo{journal}{Physics Today}}
  \textbf{\bibinfo{volume}{23}}, \bibinfo{pages}{34--41}
  (\bibinfo{year}{1970}).
\newblock 

\bibitem{Trimble1996}
\bibinfo{author}{Trimble, V.}
\newblock \bibinfo{title}{{H{\_}0: The Incredible Shrinking Constant,
  1925-1975}}.
\newblock \emph{\bibinfo{journal}{Publications of the Astronomical Society of
  the Pacific}} \textbf{\bibinfo{volume}{108}}, \bibinfo{pages}{1073}
  (\bibinfo{year}{1996}).
\newblock 

\bibitem{Burbidge1972}
\bibinfo{author}{Burbidge, G.~R.}
\newblock \bibinfo{title}{{Intergalactic Matter and Radiation (Survey
  Lecture)}}.
\newblock In \bibinfo{editor}{Evans, D.~S.}, \bibinfo{editor}{Wills, D.} \&
  \bibinfo{editor}{Wills, B.~J.} (eds.) \emph{\bibinfo{booktitle}{External
  Galaxies and Quasi-Stellar Objects, Proceedings from IAU Symposium no. 44
  held in Uppsala, Sweden, 10-14 August 1970}} (\bibinfo{publisher}{D. Reidel},
  \bibinfo{address}{Dordrecht}, \bibinfo{year}{1972}).
\newblock 

\bibitem{Rindler1967}
\bibinfo{author}{Rindler, W.}
\newblock \bibinfo{title}{{Relativistic cosmology}}.
\newblock \emph{\bibinfo{journal}{Physics Today}}
  \textbf{\bibinfo{volume}{20}}, \bibinfo{pages}{23--31}
  (\bibinfo{year}{1967}).
\newblock

\bibitem{Peebles1967}
\bibinfo{author}{Peebles, P. J.~E.} \& \bibinfo{author}{Partridge, R.~B.}
\newblock \bibinfo{title}{{Upper Limit on the Mean Mass Density due to
  Galaxies}}.
\newblock \emph{\bibinfo{journal}{The Astrophysical Journal}}
  \textbf{\bibinfo{volume}{148}}, \bibinfo{pages}{713} (\bibinfo{year}{1967}).
\newblock 

\bibitem{GottJ.R.1974}
\bibinfo{author}{{Gott, J. R.}, I.}, \bibinfo{author}{Gunn, J.~E.},
  \bibinfo{author}{Schramm, D.~N.} \& \bibinfo{author}{Tinsley, B.~M.}
\newblock \bibinfo{title}{{An Unbound Universe}}.
\newblock \emph{\bibinfo{journal}{The Astrophysical Journal}}
  \textbf{\bibinfo{volume}{194}}, \bibinfo{pages}{543} (\bibinfo{year}{1974}).
\newblock 

\bibitem{Shapiro1971}
\bibinfo{author}{Shapiro, S.~L.}
\newblock \bibinfo{title}{{The Density of Matter in the Form of Galaxies}}.
\newblock \emph{\bibinfo{journal}{The Astronomical Journal}}
  \textbf{\bibinfo{volume}{76}}, \bibinfo{pages}{291} (\bibinfo{year}{1971}).
\newblock 

\bibitem{Noonan1971}
\bibinfo{author}{Noonan, T.~W.}
\newblock \bibinfo{title}{{The Mean Cosmic Density from Galaxy Counts and Mass
  Data}}.
\newblock \emph{\bibinfo{journal}{Publications of the Astronomical Society of
  the Pacific}} \textbf{\bibinfo{volume}{83}}, \bibinfo{pages}{31}
  (\bibinfo{year}{1971}).
\newblock 

\bibitem{Ostriker1973}
\bibinfo{author}{Ostriker, J.~P.} \& \bibinfo{author}{Peebles, P. J.~E.}
\newblock \bibinfo{title}{{A Numerical Study of the Stability of Flattened
  Galaxies: or, can Cold Galaxies Survive?}}
\newblock \emph{\bibinfo{journal}{The Astrophysical Journal}}
  \textbf{\bibinfo{volume}{186}}, \bibinfo{pages}{467} (\bibinfo{year}{1973}).
\newblock 

\bibitem{Page1961}
\bibinfo{author}{Page, T.}
\newblock \bibinfo{title}{{Average Masses of the Double Galaxies}}.
\newblock In \bibinfo{editor}{{Jerzy Neyman}} (ed.)
  \emph{\bibinfo{booktitle}{Proceedings of the Fourth Berkeley Symposium on
  Mathematical Statistics and Probability, Volume 3: Contributions to
  Astronomy, Meteorology, and Physics. June 20-July 30, 1960.}},
  \bibinfo{pages}{277--306} (\bibinfo{address}{Berkeley},
  \bibinfo{year}{1961}).

\bibitem{Rood1972}
\bibinfo{author}{Peebles, P. J.~E.}
\newblock \bibinfo{title}{{Structure of the Coma Cluster of Galaxies}}.
\newblock \emph{\bibinfo{journal}{The Astronomical Journal}}
  \textbf{\bibinfo{volume}{75}}, \bibinfo{pages}{13} (\bibinfo{year}{1970}).
\newblock 

\bibitem{Page1970}
\bibinfo{author}{Page, T.}
\newblock \bibinfo{title}{{Spectral Lines and Radial Velocities of Galaxies in
  Pairs}}.
\newblock \emph{\bibinfo{journal}{The Astrophysical Journal}}
  \textbf{\bibinfo{volume}{159}}, \bibinfo{pages}{791} (\bibinfo{year}{1970}).
\newblock 

\bibitem{Janssen2002}
\bibinfo{author}{Janssen, M.}
\newblock \bibinfo{title}{{COI Stories: Explanation and Evidence in the History
  of Science}}.
\newblock \emph{\bibinfo{journal}{Perspectives on Science}}
  \textbf{\bibinfo{volume}{10}}, \bibinfo{pages}{457--522}
  (\bibinfo{year}{2002}).



\bibitem{Ellis2014}
\bibinfo{author}{Ellis, G.} \& \bibinfo{author}{Silk, J.}
\newblock \bibinfo{title}{{Scientific method: Defend the integrity of
  physics.}}
\newblock \emph{\bibinfo{journal}{Nature}} \textbf{\bibinfo{volume}{516}},
  \bibinfo{pages}{321--3} (\bibinfo{year}{2014}).
\newblock

\bibitem{Popper1963}
\bibinfo{author}{Popper, K.}
\newblock \emph{\bibinfo{title}{{Conjectures and Refutations}}}
  (\bibinfo{publisher}{Routledge}, \bibinfo{address}{London},
  \bibinfo{year}{1963}).

\bibitem{Kuhn1970}
\bibinfo{author}{Kuhn, T.~S.}
\newblock \bibinfo{title}{{Logic of Discovery or Psychology of Research?}}
\newblock In \bibinfo{editor}{Lakatos, I.} \& \bibinfo{editor}{Musgrave, A.}
  (eds.) \emph{\bibinfo{booktitle}{Criticism and the Growth of Knowledge}},
  \bibinfo{pages}{1--23} (\bibinfo{publisher}{Cambridge University Press},
  \bibinfo{address}{Cambridge}, \bibinfo{year}{1970}).

\bibitem{Abell1961}
\bibinfo{author}{Abell, G.~O.}
\newblock \bibinfo{title}{{Evidence regarding second-order clustering of
  galaxies and interactions between clusters of galaxies}}.
\newblock \emph{\bibinfo{journal}{The Astronomical Journal}}
  \textbf{\bibinfo{volume}{66}}, \bibinfo{pages}{607} (\bibinfo{year}{1961}).
\newblock

\bibitem{Longair1971}
\bibinfo{author}{Longair, M.~S.}
\newblock \bibinfo{title}{{Observational cosmology}}.
\newblock \emph{\bibinfo{journal}{Reports on Progress in Physics}}
  \textbf{\bibinfo{volume}{34}}, \bibinfo{pages}{306} (\bibinfo{year}{1971}).
\newblock

\end{thebibliography}


}
\begin{addendum}
  \setlength\itemsep{0em}

 \item We warmly thank Virginia Trimble for her elaborate comments and suggestions, and all interviewees for their interest and cooperation. The interviews in this article have been made possible by a grant-in-aid from the Friends of the Center for History of Physics, American Institute of Physics, and the kind support of the Department of Astrophysical Sciences, Princeton University. This work is partly financed by the Netherlands Organisation for Scientific Research (NWO; project number SPI 63-260). G.B. acknowledges support from the European Research Council through the ERC starting grant {\it WIMPs Kairos}.
 \item[Additional Information] The transcripts of the interviews are archived at the Niels Bohr Library \& Archive, American Institute of Physics.
 \item[Author Contributions] J.d.S. has conducted the historical research, conducted the interviews, and prepared the manuscript. G.B. and J.v.D. defined the project, supervised the research, gave technical and conceptual advice, and contributed to the writing of the manuscript.
 
  \item[Correspondence] Correspondence  should be addressed to j.g.deswart@uva.nl.

\end{addendum}
\vspace{12pt}

\end{document}